\documentclass{Rinton-P9x6}
\usepackage{latexsym}
\newcommand \vek[1]{\mathbf{#1}}

\begin{document}
\title{Calculating Vacuum Energies in \\Quantum Field Theory}
\author{Markus Quandt}
\address{Institute for Theoretical Physics\\
University of T\"ubingen\\
D-72076 T\"ubingen, Germany\\
}
\maketitle


\abstracts{A new approach to generalised Casimir type of problems is derived 
within the context of renormalisable quantum field theory (QFT). We study
the simplest case of a massive fluctuating boson field coupled to a 
time-independent background potential. We use analytic properties of 
scattering data to compute the relevant Green's functions at imaginary 
momenta, which in turn yields a simple and efficient method to compute
(one-loop) vacuum energy densities in QFT. Renormalisation is easily performed 
in the perturbative sector by identifying low order Feynman diagrams with 
the first few Born approximation to the Green's function. Numerical 
examples illustrate the efficiency of our approach.
}


\section{Introduction}

In this talk, I report on the technical aspects of a 
novel quantum field theory approach to generalised Casimir problems. 
The work presented here was done in collaboration with N.~Graham, 
R.~L.~Jaffe, V.~Khemani, M.~Scandurra, O.~Schr\"oder and H.~Weigel; 
for a more complete exposition, see ref.\cite{density}.   

The traditional approach to Casimir systems imposes boundary conditions 
on a quantum field \emph{ab initio}\cite{traditional}. In reality, however, 
Casimir forces arise from \emph{interactions} between the quantum field and
matter, and no such interaction is strong enough to enforce a boundary 
condition on \emph{all} frequencies of the fluctuating field. 
To study whether an idealised boundary condition limit exists in which 
physical quantities become independent of the regularisations 
involved, we have recently proposed\cite{letter} to embed the Casimir 
calculation in a quantum field theory (QFT). The external condition is 
then replaced by a renormalisable interaction with a smooth background 
field that imposes the boundary condition in a certain limit.
QFT renormalisation is the only sensible way to discuss and 
eventually remove UV divergences. Any remaining infinities in our approach 
indicate that the quantities under consideration depend in detail on the 
physical UV cutoff provided by the material. In ref.\cite{dirichlet}
we argued that the traditional approach is perfectly valid for some 
quantities (e.g.~forces between rigid bodies) while it fails for 
others (e.g.~the Casimir stress on isolated surfaces).

Since the boundary condition limit in the QFT approach involves arbitrarily
strong and sharply peaked background fields, the standard tools for 
the calculation of effective energies (or functional determinants)
cannot be employed: Perturbation theory fails because of the strong 
coupling, while a derivative expansion is ruled out by the high Fourier 
modes in the background field. Thus, we had to develop general new 
methods to compute renormalized one-loop quantum energies and energy 
densities.\footnote{It should be mentioned that the \emph{world-line 
formalism}\cite{worldline} provides an alternative way for an exact 
(numerical) computation of 1-loop quantum energies. It is even applicable
in cases where there is not enough radial symmetry for our method to work.} 
These methods are 
important in their own right and have a much broader application than the 
proper Casimir problem studied in ref.\cite{dirichlet}; for details the reader 
is referred to refs.\cite{applic}. In the present talk, I will concentrate on 
the derivation of our approach from conventional quantum field theory,  
which is presented in section \ref{sec:2}. Computational techniques for 
Green's functions and renormalisation are discussed in section \ref{sec:3}. 
Section \ref{sec:4} presents a numerical example to demonstrate the 
efficiency of our method. I conclude with a brief summary and outlook 
on future directions.


\section{The Method}
\label{sec:2}

For simplicity, I will restrict myself to the case 
of a massive scalar boson $\phi$ interacting with a time-independent scalar
background field $\sigma(\vek{x})$ through the coupling 
$\mathcal{L}_{\rm int} = - \frac{\lambda}{2} \phi^2 \sigma(\vek{x})$. This 
model is renormalisable (in $n\le 3$ space dimensions) and can be used 
as a QFT implementation of \emph{Dirichlet b.c.}~on a 
surface $\mathcal{S}$,\cite{dirichlet} if we let the background $\sigma$ 
be strong and sharply concentrated on $\mathcal{S}$. In general, 
our method requires that the background is sufficiently short ranged so
that a conventional scattering theory can be defined. More importantly, we
require enough (spherical) symmetry to separate the scattering problem 
into partial waves; we shall consequently assume that $\sigma(\vek{x})$ 
only depends on $r=|\vek{x}|$.
The vacuum energy density in a spherical shell of radius $r$ is then   
\be
\epsilon(r) \equiv \frac{2 \pi^{n/2}}{\Gamma(n/2)}\,r^{n-1} 
\left\langle 
\Omega \left| \frac{1}{2}\left[\dot{\phi}^2 + 
(\nabla \phi)^2 + m^2 \phi^2 + \sigma(r)\phi^2\right]
\right| \Omega  \right\rangle_{\rm ren}
\label{eq1}
\ee
where $|\Omega\rangle$ is the interacting vacuum
and the operator in the vev is just the 
symmetric energy density operator $\hat{T}_{00}$ for our model.
To further evaluate eq.~(\ref{eq1}), we first perform a partial wave 
decomposition of the field operator,
$
\phi(t,\vek{x}) = \sum_{\{\ell\}} \phi_\ell(t,r)\,\mathcal{Y}_{\{\ell\}}(
\hat{x})
$
where $\mathcal{Y}_{\{\ell\}}(\hat{x})$ are the $n$-dimensional spherical 
harmonics and $\{\ell\}$ refers to the set of all angular quantum numbers in 
$n$ space dimensions. Finally, we make a \emph{Fock decomposition} for 
the radial operators,
\be
\phi_\ell(t,r) = r^{\frac{1-n}{2}} \int_0^\infty \,\frac{d k}
{\sqrt{\pi \omega}}\,\left[\psi_\ell(k,r)\,e^{- i \omega t} a_\ell(k) + 
\psi_\ell^\ast(k,r)\,e^{i \omega t} a_\ell^\dagger(k) \right] + 
\mbox{b.s.}
\label{eq2}
\ee
with $\omega = \sqrt{m^2 + k^2}$. The wave functions $\psi_\ell(k,r)$ 
in eq.~(\ref{eq2}) are a complete set of scattering and bound 
state\footnote{The term "b.s." in eq.~(\ref{eq2}) indicates a similar 
contribution from the bound states of energy $\omega_j = 
\sqrt{m^2 - \kappa_j^2}$, which are required for the completness of 
the function system.} solutions to the fully interacting 
field equations. They can be ortho-normalised such that the expansion
coefficients $a_\ell(k)$ obey the standard commutation relations
$[a_\ell(k),a^\dagger_{\ell'}(k')] = \delta(k-k')\delta_{\ell\ell'}$,
(and similarly for the b.s.) which allows the usual particle interpretation 
(e.g.~$a_\ell^\dagger(k)$ creates a $\psi_\ell$-mode acting on 
$|\Omega\rangle$). 

Inserting our decomposition back into eq.~(\ref{eq1}), the result is best
expressed in terms of the \emph{Green's function}
\be
G_\ell(r,r';k) = - \frac{2}{\pi} \int_0^\infty dq\,\frac{\psi_\ell^\ast(q,r)
\psi_\ell(q,r')}{(k+i\epsilon)^2 - q^2} - \sum_j \frac{\psi_{\ell,j}(r)
\psi_{\ell,j}(r')}{k^2 + \kappa_j^2}
\label{eq3}
\ee
which is meromorphic in the upper complex $k$-plane with simple poles at
the bound state momenta $k=i \kappa_j$ (in the next
section, I will sketch an efficient method to compute $G_\ell$). The 
Green's function is intimately related to the density of states. In fact,
defining a \emph{local spectral density}
\be
\rho_\ell(k,r) \equiv - i k\,G_\ell(r,r;k)\,,\qquad\quad
\mathsf{Im}\,k \ge  0
\label{eq4}
\ee
in the upper $k$-plane, we easily recover the local density of states 
for real $k$, 
\[
\mathsf{Re}\,\rho_\ell(k,r) = \mathsf{Im}\,\{k\,G_\ell(r,r;k)\}
= \psi_\ell^\ast(k,r)\psi_\ell(k,r)\,. 
\]
The energy density, eq.~(\ref{eq1}), can be expressed (up to a 
total derivative) as an integral of the single particle energies $\omega(k)$
over all real $k$, weighted by the local density of states in each channel
$\ell$ (plus a contribution from the bound states $\omega_j$).  
For practical purposes, however, it is much more convenient to 
rotate the integration contour to the upper complex $k$-plane. We get 
three contributions: (1) The residues from the bound state poles on the 
imaginary axis cancel the explicit bound state contribution in $G_\ell$
exactly,\cite{Bordag} (2) the discontinuity of the square root cut in 
the single particle scattering energies $\omega = \sqrt{m^2 + k^2}$ 
yields an integral along the imaginary axis, $k= i t$, $t\in[m,\infty]$, 
and (3) the big semicircle at infinity gives no contribution if the 
integrand falls off fast enough for $|k|\to\infty$. To ensure (3), we 
have to improve the decay of the local spectral density 
eq.~(\ref{eq4}) at large $|k|$. Since the \emph{Born series} to the 
Green's function converges at large $|k|$, it is sufficient to subtract 
a few low order Born approximations. We use the notation
\be
\left[\rho_\ell(k,r)\right]_N \equiv \rho_\ell(k,r) - 
\rho_\ell^{(0)}(k,r) - \cdots - \rho_\ell^{(N)}(k,r)
\label{eq5}
\ee
for the $N$ times Born subtracted density (and similarly for the 
Greens function). Formally, the Born series is an expansion of
$G_\ell$ in powers of the interaction, i.e.~the coupling strength 
$\lambda$. When used in the expression for the energy density 
$\epsilon(r)$, the Born terms should thus correspond to 
the usual perturbative \emph{Feynman series}. Notice that we need 
a cutoff at this point since the low order Feynman diagrams and 
Born approximations to $G_\ell$ are precisely the UV divergent parts of 
the calculation. Using dimensional regularisation, the identification 
of Born and Feynman series has been established rigorously at least 
for the lowest orders\cite{applic}. The final result for the 
energy density in a spherical shell of radius $r$ is 
then\footnote{Since the vacuum bubble diagram $\epsilon_{\rm FD}^{(0)}$ is 
not inserted back, eq.~(\ref{eq6}) really represents the
\emph{change} in the vacuum energy due to the interaction.}
\begin{eqnarray}
\epsilon(r) &=& - \sum_\ell N_\ell\int_m^\infty \frac{d t}{\pi}\,
\sqrt{t^2 - m^2}\,\left[1 - \frac{1}{4 (t^2 - m^2)} D_r\right]
\,\left[\rho_\ell(it,r)\right]_N + \nonumber\\
&& {}\qquad + \sum_{i=1}^{N} \epsilon_{\rm FD}^{(i)} + 
\epsilon_{\rm CT}(r)\,.
\label{eq6} 
\end{eqnarray}
Here, $N_\ell$ is the multiplicity in the channel $\ell$, 
$D_r = \partial_r(\partial_r - (n-1)/r)$ is a total derivative operator
and $\epsilon_{\rm CT}$ denotes the \emph{counter terms} necessary to 
renormalise the $N$ low order Feynman diagrams $\epsilon_{\rm FD}^{(i)}$.
The perturbative renormalisation of $\epsilon_{\rm FD}$ is standard and 
briefly sketched in the next section, where I also present an efficient 
method to evaluate the local density $\rho_\ell$ and its Born series. 
It should be emphasised that eq.~(\ref{eq6}) yields a finite 
renormalised vacuum energy density for any smooth background $\sigma(r)$. 
Renormalisation group arguments prove that the 
final result is cutoff- and scheme-independent. 

To compute the \emph{quantum energy} $E_q[\sigma]$ we can simply integrate
eq.~(\ref{eq6}) over all radii. Using the formula\cite{density} 
\be
2 \int _0^\infty dr\,[\rho_\ell(k,r)]_0 = i \frac{d}{d k} 
\ln F_\ell(k)
\ee
valid in the upper complex $k$-plane, we can relate the density of states
in $k$-space to the log-derivative of the \emph{Jost function}
$F_\ell(k)$. Rotating back to the real axis from above, this turns into 
the well-known expression
\be
[\rho_\ell(k)]_0 = \rho_\ell(k) - \rho_\ell^{(0)}(k) = 
\frac{1}{\pi}\,\frac{d\delta_\ell}{d k}
\ee
for the (change in the) density of states in terms of the 
\emph{phase shift} $\delta_\ell$. Thus we have the quantum energy
($\equiv$ change in the zero-point energy due to the background $\sigma$)
\be
E_q[\sigma] = \sum_\ell N_\ell \int_m^\infty \frac{d t}{2\pi}\,
\frac{t}{\sqrt{t^2-m^2}}\left[\nu_\ell(t)\right]_N + 
\sum_{i=1}^N E_{\rm FD}^{(i)} + E_{\rm CT}
\label{eq7}
\ee
where $\nu_\ell(t) = \ln F_\ell(it)$ is the log of the Jost function 
on the imaginary axis. 

The continuation of eq.~(\ref{eq7}) back to the real axis has been 
used extensively in the past\cite{applic}. Though eq.~(\ref{eq7})
is superior from a computational point of view, the formula on the real
axis has the merit of a clear physical interpretation in terms of 
one-particle states and densities.
In the next section, I will briefly discuss the methods necessary to 
turn eq.~(\ref{eq6}) and (\ref{eq7}) into efficient computational tools.


\section{Computational Techniques and Renormalisation}
\label{sec:3}

The Feynman series for the energy density is obtained from the
usual perturbative expansion\footnote{For the total energy eq.~(\ref{eq7}),
the Feynman diagrams are more easily computed from the functional 
determinant representation $E_q = \frac{1}{2T}\ln\det(-\Box + m^2 + \sigma)$ 
in euclidean space ($T$ is the euclidean time interval).}
\be
\langle 0 | \hat{T}_{00}(x) | 0 \rangle = 
\frac{1}{2 i} \mathrm{Tr}\,\left[\hat{T}_x 
( - \partial^2 - m^2 - \sigma)^{-1}\right]
\label{eq8}
\ee
where $\hat{T}_x$ is the coordinate space operator corresponding to 
the insertion of the energy density (\ref{eq1}) at the space-time 
point $x$. It has pieces of order $\sigma^0$ and $\sigma^1$ which 
read, in momentum space, 
\begin{eqnarray}
\langle k'|\hat{T}_x^{(0)}|k\rangle &=& 
e^{i(k'-k)x}\,\left[k'^0 k^0 + \vek{k}'\cdot\vek{k} + m^2\right]\\\nonumber
\langle k'|\hat{T}_x^{(1)}|k\rangle &=& \sigma(x) e^{i (k'-k)x}\,. 
\end{eqnarray}
The Feynman series consists of all graphs with a single $\phi$-loop
and arbitrary insertions of $\hat{T}_x^{(0)}$, $\hat{T}_x^{(1)}$
and $\sigma(x)$ (from expanding the propagator in eq.~(\ref{eq8})). 
To order $\sigma^1$, for instance, we have two diagrams:
(a) a single insertion of $\hat{T}_x^{(1)}$ and (b) one insertion of 
$\sigma$ and one insertion of $\hat{T}_x^{(0)}$. Only diagram (a)
(the \emph{tadpole graph}) is divergent,
\be
\frac{1}{2 i} \mathrm{Tr}\,\left[\hat{T}^{(1)}_x 
(-\partial^2 - m^2)^{-1}\right] = \frac{i}{2} \sigma(\vek{x}) 
\int \frac{d^d k}{(2\pi)^d}\,\frac{1}{k^2 - m^2}\,. 
\ee
It may easily be renormalised (and even cancelled completely) by a 
counter term proportional to $\sigma(\vek{x})$. For the present model 
the tadpole counter term $c_1 \sigma$ and one Born subtraction $N=1$ are
in fact sufficient for space dimension up to $n=2$. In $n=3$, an additional 
mass counter term $c_2 \sigma^2$ is required and we have to perform $N=2$ 
Born subtractions. The coefficient $c_2$ may be fixed by a specific choice of 
renormalisation condition but the renormalisation group ensures that physical 
quantities are scheme independent. 

More generally, the counter terms are \emph{local monomials} in $\sigma$ and 
$\partial\sigma$. It is then immediately clear that renormalisation will only 
affect the energy density on the \emph{support} of the counter terms.
For a Casimir type of calculation, the background becomes sharply 
concentrated on the b.c.~surfaces and the energy density away from the 
surfaces cannot be affected by the counterterms; it must thus be finite.
The divergence of the energy density as one approaches Casimir surfaces 
has in fact been known for a long time.\cite{candelas}

The final ingredient in our main formula, eq.~(\ref{eq6}), is the 
appropriate Green's function, which may be expressed as a product of
Jost ($f_\ell$) and regular ($\phi_\ell$) solution to the scattering 
problem,
\[
G_\ell(r,r';k) = (-k)^{\ell + (n - 3)/2}\,\,\frac{\phi_\ell(k,r_{<})\,
f_\ell(k,r_{>})}{F_\ell(k)}\,.
\]
In order to evaluate this formula on the imaginary $k$-axis, it is important 
to cancel all oscillating pieces in the numerator, which would 
otherwise become numerically intractable in the upper complex $k$-plane. 
We cancel the free Jost solution $w_\ell(kr)$ (a modified Hankel 
function) explicitly by the ansatz $f_\ell = w_\ell \,g_\ell$ and 
$\phi_\ell \sim h_\ell / w_\ell$. The result is a representation of the 
form\footnote{The $s$-wave in $n=2,3$ requires a slightly different
treatment\cite{density}.}
\be
G_\ell(r,r;k) = \frac{h_\ell(k,r) g_\ell(k,r)}
{(2\ell - 2 + n) g_\ell(k,0)} 
\ee
in terms of two new functions $g_\ell$ and $h_\ell$. Since both the Jost 
and regular solution obey the full radial field equation with certain 
boundary conditions, it is easy to derive ODE's for the relevant  
functions $g_\ell$ and $h_\ell$. The explicit expressions are somewhat 
lengthy and the reader is referred to ref.\cite{density} for details.
A few general remarks are in order. By introducing \emph{two} 
functions with boundary conditions at $r=0$ (for $h_\ell$) and at 
$r=\infty$ (for $g_\ell$) we can avoid expensive shooting methods for 
the Green's function $G_\ell$. The solutions $g_\ell$ and $h_\ell$ 
can be shown to be smooth \emph{bounded} functions of $r$ in the 
upper complex $k$-plane. The ODEs may, in fact, be solved directly 
on the imaginary axis $k=it$, where the solutions are 
\emph{real} and easily amenable to numerical computations.
The Born approximations are most conveniently computed by iterating the 
ODEs for $g_\ell$ and $h_\ell$ according to an expansion of $G_\ell$ 
in powers of the interaction $\sigma$. 

Finally, a similar ansatz known as the \emph{variable phase approach} 
can be used for the log of the Jost function required in the quantum 
energy formula, eq.~(\ref{eq7}). Writing $g_\ell(it,r) = e^{\beta_\ell(t,r)}$ 
with a \emph{real} function $\beta_\ell$, we find the relevant function
$\nu_\ell(t) \equiv \beta(t,0)$ by integrating the ODE
\be
- \beta_\ell''(t,r) - [\beta_\ell'(t,r)]^2 + 2 t \,\xi_\ell(tr)
\,\beta_\ell'(t,r) + \sigma(r) = 0
\ee 
inwards, starting at $r\to\infty$ with $\beta_\ell(t,r) = 
\beta_\ell'(t,r) = 0$ (Born approximations follow again by iteration).
The coefficient function $\xi_\ell$ is just the log-derivative of the 
free Jost function evaluated on the imaginary axis; for details
see ref.\cite{density}.

\begin{figure}[t]
\begin{minipage}{5cm}
 \epsfxsize=5cm \epsfbox{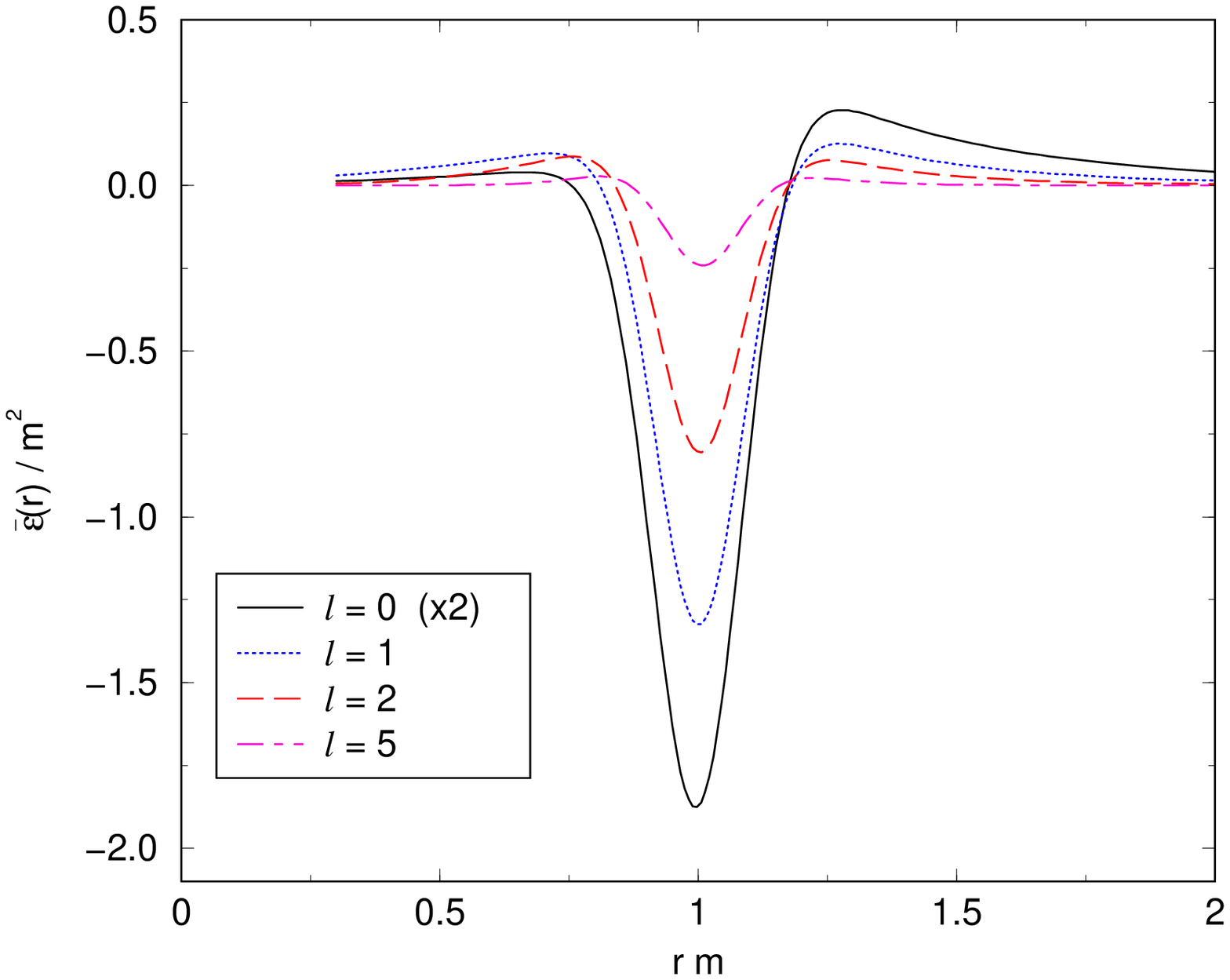}
\end{minipage}
\hfill
\begin{minipage}{5cm}
 \epsfxsize=5cm \epsfbox{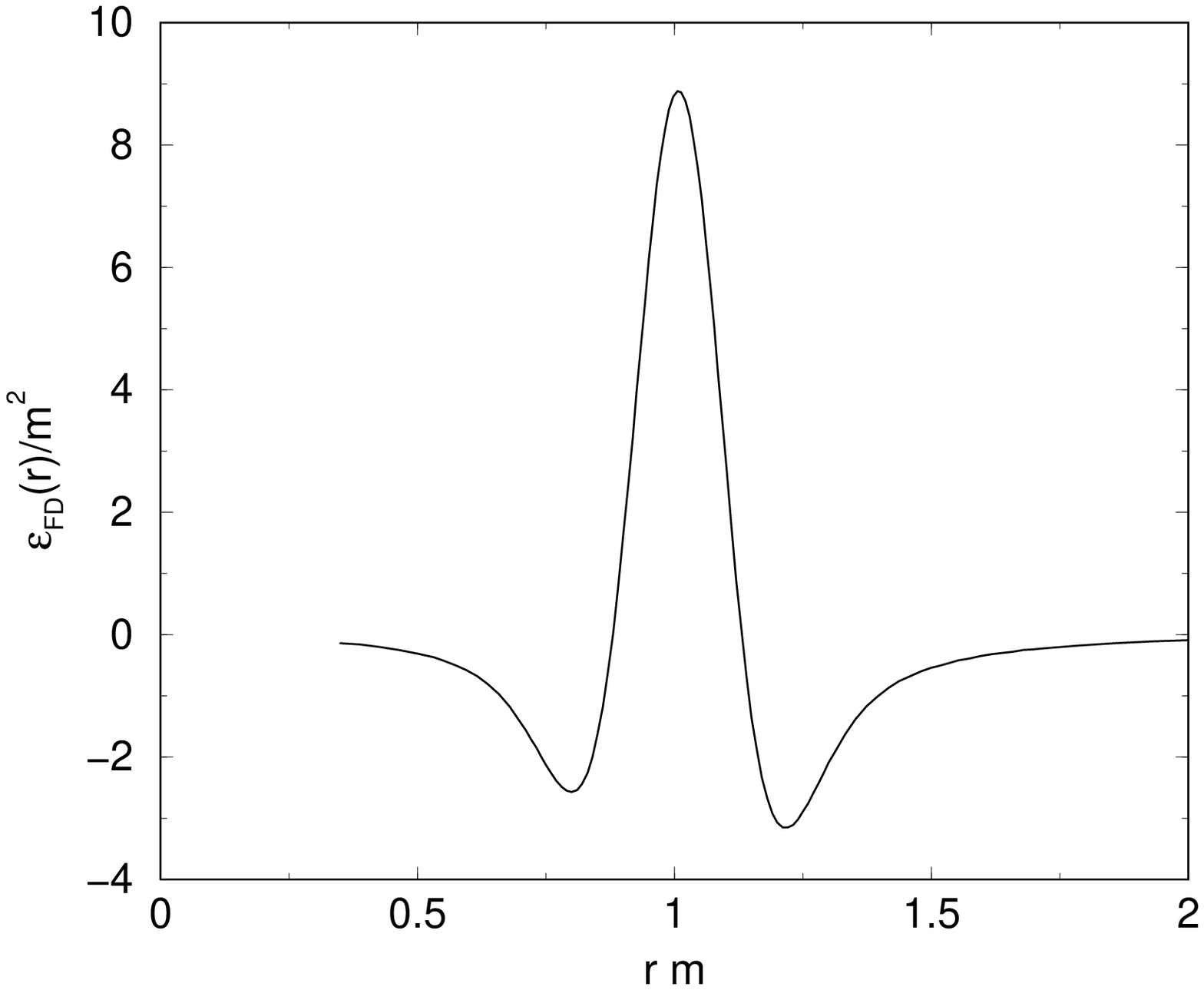}
\end{minipage} 
\caption{Left panel: Contributions of various angular momentum channels to 
the quantum energy density of the Gaussian ring. The first order Feynman 
diagram depicted in the right panel was excluded on the left. All quantities
are measured in units of the boson mass $m$.} 
\label{fig:1}
\end{figure}


\section{A Numerical Example}
\label{sec:4}

As a numerical example for the efficiency of our method, I 
discuss a background field $\sigma(r)$ which takes the profile of a 
Gaussian ring in $n=2$ space dimensions,
$
\sigma(r) = A \exp[- \frac{(r-a)^2}{2 w^2} ]\,.
$ 
In the limit where the width $w$ tends to zero (and the strength to infinity) 
this background can be used to study the Casimir stress on a Dirichlet circle 
in $n=2$ dimensions. Here, I will employ it as a numerical example of 
how our method handles sharply peaked backgrounds. 

The left panel of figure \ref{fig:1} shows the contributions of various 
angular momentum channels to the quantum energy density. These terms 
correspond to the once Born subtracted $t$-integral in eq.~(\ref{eq6}),
i.e.~they do not include the $1^{\rm st}$ order Feynman diagram\footnote{Note
that the tadpole diagram has been cancelled completely by the counterterm,
but there is still a second (finite) contribution of order $\sigma^1$
as explained in the last section.}
depicted in the right panel of fig.~\ref{fig:1}. From the scales at the 
axes, it is clear that the first order diagram dominates the energy density 
(i.e.~perturbation theory is accurate) even at relatively large couplings 
(fig.~\ref{fig:1} corresponds to $\lambda/m = 3.0$). 

\begin{figure}[t]
\centerline{\begin{minipage}{6.5cm}
 \epsfxsize=6.5cm \epsfbox{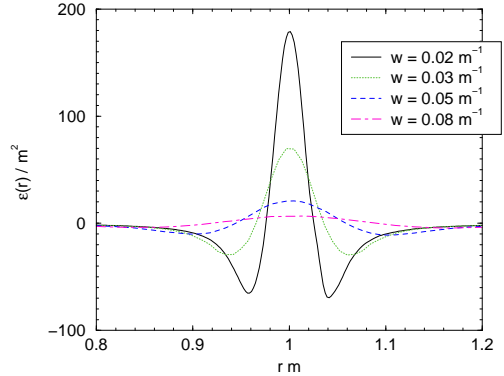}
\end{minipage}}
\caption{Quantum energy density for Gaussian rings of various 
widths $w$.}  
\label{fig:2}
\end{figure}

Figure \ref{fig:2} shows the complete quantum energy for Gaussian circles
of various widths $w$. As can be seen from the height of the central peak, 
the density \emph{on the surface} diverges in the sharp limit $w\to 0$ 
even after renormalisation. Our method allows to study this limit 
numerically by going to very small widths for which traditional 
approximation schemes fail. 


\section{Conclusion}
\label{sec:5}

In this talk, I have presented a new approach to the computation 
of 1-loop vacuum energies and energy densities in quantum field 
theory. Starting from a conventional Fock decomposition or, 
alternatively, from a Green's function approach, we 
identify the potential UV divergent contributions to the quantum 
energy with the low order Born approximations to the Green's 
function. We subtract these contributions and add them back  
as Feynman diagrams which allows for a conventional and 
scheme independent renormalisation using standard counter terms.
The resulting formulae for the Casimir energy (density) are 
finite functionals of smooth background fields and easily
amenable to numerical treatment. As an example, I have shown 
results for the Casimir energy density of a Gaussian circle
in $n=2$ space dimensions.  

Our method has many interesting applications. Among the topics currently 
under investigation are magnetic vortices (which play a role
in the confinement properties of gauge theories) and $Z$-strings 
in electroweak theory, which bind fermions efficiently and may 
thus be of relevance in scenarios of baryogenesis. 

\end{document}